\documentclass[%
 aip,
 jap,%
 amsmath,amssymb,
preprint,
]{revtex4-1}

\usepackage{graphicx}
\usepackage{color}
\usepackage{dcolumn}
\usepackage{bm}



\begin{document}


\title{Parameter space for thermal spin-transfer torque}

\author{J.~C.~Leutenantsmeyer}
\email{jleuten@gwdg.de}
\author{M.~Walter}
\author{V.~Zbarsky}
\author{M.~M\"unzenberg}
\affiliation{I. Physikalisches Institut, Georg-August-Universit\"at G\"ottingen, Friedrich-Hund-Platz 1, 37077 G\"ottingen,~Germany}%

\author{R. Gareev}
\affiliation{Institut f\"ur Experimentelle und Angewandte Physik, Universit\"at Regensburg, Universit\"atsstrasse 31, 93040 Regensburg,~Germany}

\author{K. Rott}
\author{A.~Thomas}
\author{G. Reiss}
\affiliation{Thin Films and Physics of Nanostructures, Universit\"at Bielefeld \\  
Universit\"atsstrasse 25 \\ 33615 Bielefeld,~Germany}

\author{P.~Peretzki}
\author{H.~Schuhmann}
\author{M.~Seibt}
\affiliation{IV. Physikalisches Institut, Georg-August-Universit\"at G\"ottingen, Friedrich-Hund-Platz 1, 37077 G\"ottingen,~Germany}%

\author{M. Czerner}
\author{C. Heiliger}
\affiliation{I. Physikalisches Institut, Universit\"at Gie\ss en, Heinrich-Buff-Ring 16, 35392 Gie\ss en, Germany}%

\date{\today}
             
\begin{abstract}
Thermal spin-transfer torque describes the manipulation of the magnetization by the application of a heat flow. The effect has been calculated theoretically by Jia et al. in 2011. It is found to require large temperature gradients in the order of Kelvins across an ultra thin MgO barrier. In this paper, we present results on the fabrication and the characterization of magnetic tunnel junctions with 3 monolayer thin MgO barriers. The quality of the interfaces at different growth conditions is studied quantitatively via high-resolution transmission electron microscopy imaging. We demonstrate tunneling magneto resistance ratios of up to 55\% to 64\% for 3 to 4 monolayer barrier thickness. Magnetic tunnel junctions with perpendicular magnetization anisotropy show spin-transfer torque switching  with a critical current of 0.2\,MA/cm$^2$. The thermally generated torque is calculated ab initio using the Korringa-Kohn-Rostoker and non-equilibrium Green's function method. Temperature gradients generated from 
femtosecond laser pulses were simulated using COMSOL, revealing gradients of 20\,K enabling thermal spin-transfer-torque switching.
\end{abstract}

\keywords{magnetic tunnel junctions, thermal spin-transfer torque, ultra thin MgO barrier}
\maketitle

\section{Introduction}
Spintronic devices provide excellent opportunities for data storage applications.
Today, magnetic random access memory (MRAM) has shown several advantages to 
conventional RAM. Besides faster access times also higher storage density, 
lower power consumption and non-volatility are obtained.\cite{Yuasa2007}

Spincaloric effects in magnetic tunnel junctions (MTJs) may provide a great way 
of using excess heat for storage devices.\cite{Bauer2012,LeBreton2011} Utilizing effects such as the tunneling 
magneto-Seebeck (TMS) effect, the energy efficiency of memory 
will be enhanced and mark a next step towards a greener information technology. 
This effect was recently predicted theoretically\cite{Czerner2011} and observed 
experimentally.\cite{Walter2011,Liebing2011,Lin2012}

In addition to the TMS effect, thermal spin-transfer torque (T-STT), has been proposed by Slonczewski\cite{Slonczewski2010} in 2010 and calculated by Jia et al.\cite{Jia2011} in 2011. Corresponding to the conventional spin-transfer torque effect\cite{Slonczewski1996,Berger1996}, a spin-polarized tunneling current is used to switch the state of an MTJ. In case of T-STT, this current is generated from the thermally excited electron transport across the tunneling barrier.

The theoretical considerations of T-STT lead to experimental challenges for the sample preparation and the experiments: 
3 monolayer (ML, 0.63\,nm) thin MgO barriers as well as large temperature gradients in the order of Kelvins across this barrier are required.\cite{Jia2011} Here, we demonstrate that all requirements can be fulfilled. 

\section{Sample preparation}
The samples are prepared on thermally oxidized silicon substrates in a ultra high vacuum (UHV) chamber. 
The MTJ stack consists of Ta (10~nm) / Co-Fe-B (2.5) / MgO (0.63--2.1) / Co-Fe-B 
(5.4) / Ta (5.0) / Ru (3.0). The thickness of the MgO is varied from 3~ML (0.63\,nm) up to 10~ML (2.1\,nm). For MTJs with perpendicular magnetization anisotropy (PMA) the thickness of the Co-Fe-B electrodes is reduced to 1.0 and 1.2\,nm.

Tantalum and Co-Fe-B are deposited in a magnetron sputter chamber, while the MgO barrier and the ruthenium capping layer  
are e-beam evaporated in an adjacent chamber without breaking the vacuum. The layer stack is annealed ex-situ to crystallize the Co-Fe-B electrodes in a solid state epitaxy process. 

After annealing, the samples are patterned using standard UV- and electron-beam 
litho\-gra\-phy and structured by argon ion milling down to diameters of 150\,nm. 


\section{Growth of ultra thin barriers}
Growing ultra thin MgO barriers introduces experimental challenges, because crystalline MgO barriers give rise to large TMR ratios due to their spin-filter effect.\cite{Parkin2004,Yuasa2004}
This effect is decreased for thin barriers.\cite{Butler2001}
While thick barriers of 10\,ML thickness can yield TMR ratios of over 
600\% at room temperature, this values is strongly decreased for thin tunnel barriers.\cite{Ikeda2008} 
A decrease of the spin polarization from additional d-like contributions was reported for alumina junctions as well.\cite{MiMueMo2011}

MgO deposited on amorphous Co-Fe-B usually crystallizes in the (001) direction if five or more ML 
are deposited. The Co-Fe-B crystallizes during annealing, while MgO acts as a template 
for the solid state epitaxy process\cite{Yuasa2005} and long range order 
on the atomic scale is induced at Co-Fe-B/MgO interfaces.\cite{Eilers2009} 
For ultra thin barriers, this effect is not present and thus the 
interface quality is reduced. A successful preparation of a TMR device below 5 ML of MgO
has to focus on a crystalline growth below that critical thickness. 

\begin{figure}[htbp]
\centerline{\includegraphics[width=0.7\linewidth]{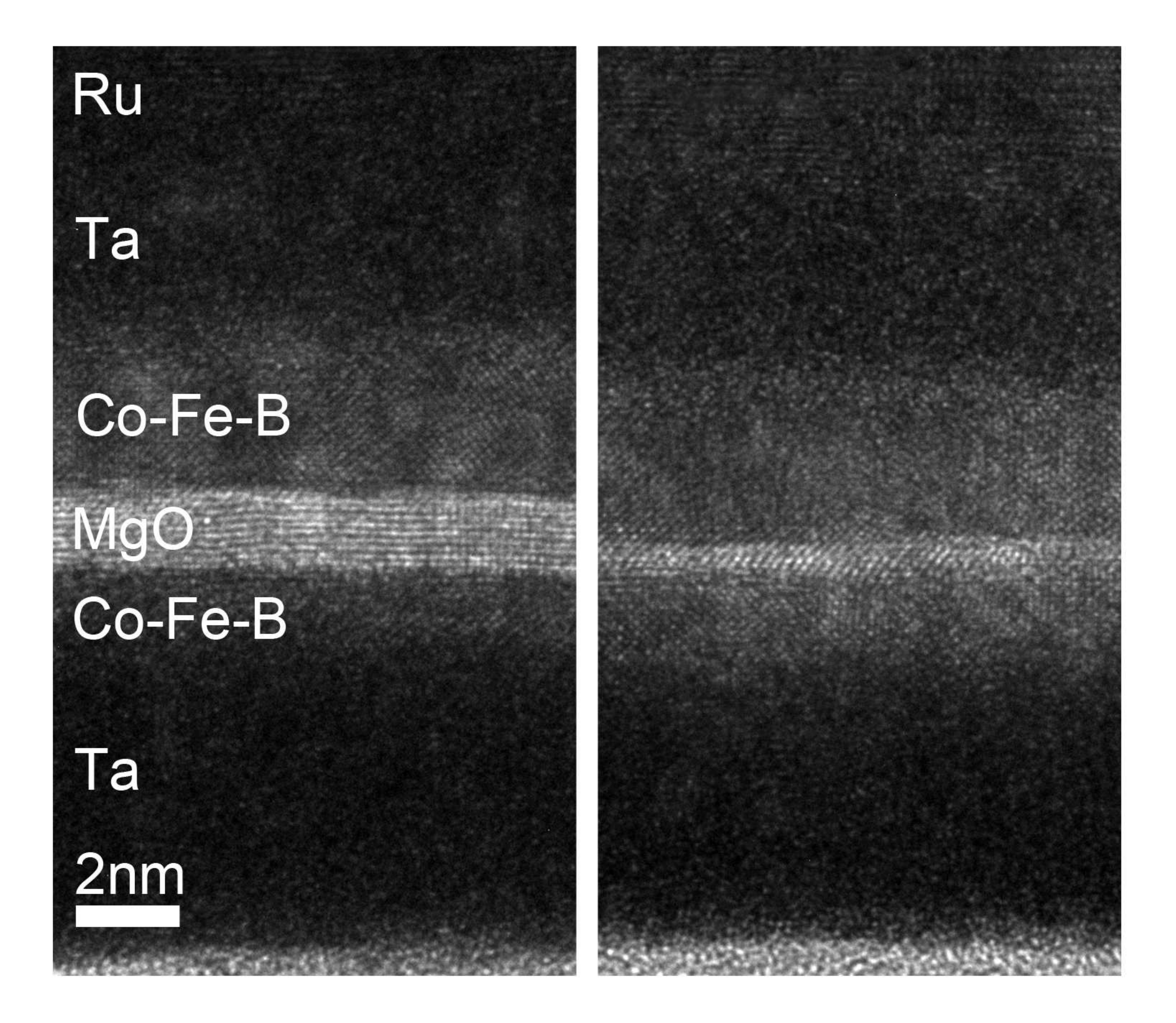}}
\caption{HRTEM images of a thick 10~ML (left) and a heated 3 ML MgO barrier 
(right). The IQR values are $(5.6~\pm~1.5)^\circ$ (10~ML) and ($6.7~\pm~0.8)^\circ$ (3~ML).}
\label{vergleich_dick_duenn}
\end{figure}

The effect of the deposition rate and influence of substrate temperature on the shift of the critical MgO 
crystallization thickness has been discussed by Kurt et al.\ and Isogami et al. A threshold value 
of 5\,pm/s MgO deposition rate is reported to obtain high quality tunnel barriers. Also, 
infrared heating to 300$^\circ$C is reported to enhance the TMR ratio of 4\,ML 
thick barriers reaching values of more than 200\%.\cite{Isogami2008,Kurt2010} The interfaces of our junctions exhibit
their best properties at deposition rates of 1\,pm/s. Furthermore, we investigated the effect of sample annealing between 100 and 350$^\circ$C during e-beam evaporation of the MgO barrier.


\section{Quantitative HRTEM analysis of MgO barriers}
Transmission electron microscopy (TEM) was conducted at a Philips CM200 FEG UT under 200~kV, 
with cross-section specimens prepared by a 
FEI Nova NanoLab 600 Focused Ion Beam under 30~kV with a final polishing step at 5~kV.
High resolution TEM images of the MgO barrier were Fourier-filtered around the MgO 
growth direction and MgO reciprocal lattice constant, which results in the MgO layer 
being represented by distorted lines arising from its crystal texture. This effect is 
used to estimate the degree of texturing of the layer by transforming the image into a two-dimensional line parameter space using the Hough transform.\cite{Duda1972} The texture leads to intensity peaks with finite distributions in the 
transformed image. The width of these distributions is estimated by their inter-quartile 
range (IQR) for the parameter pairs of several lines, whose average value is used as the 
approximation for one image. 

\begin{figure}[htbp]
\centerline{\includegraphics[width=0.7\linewidth]{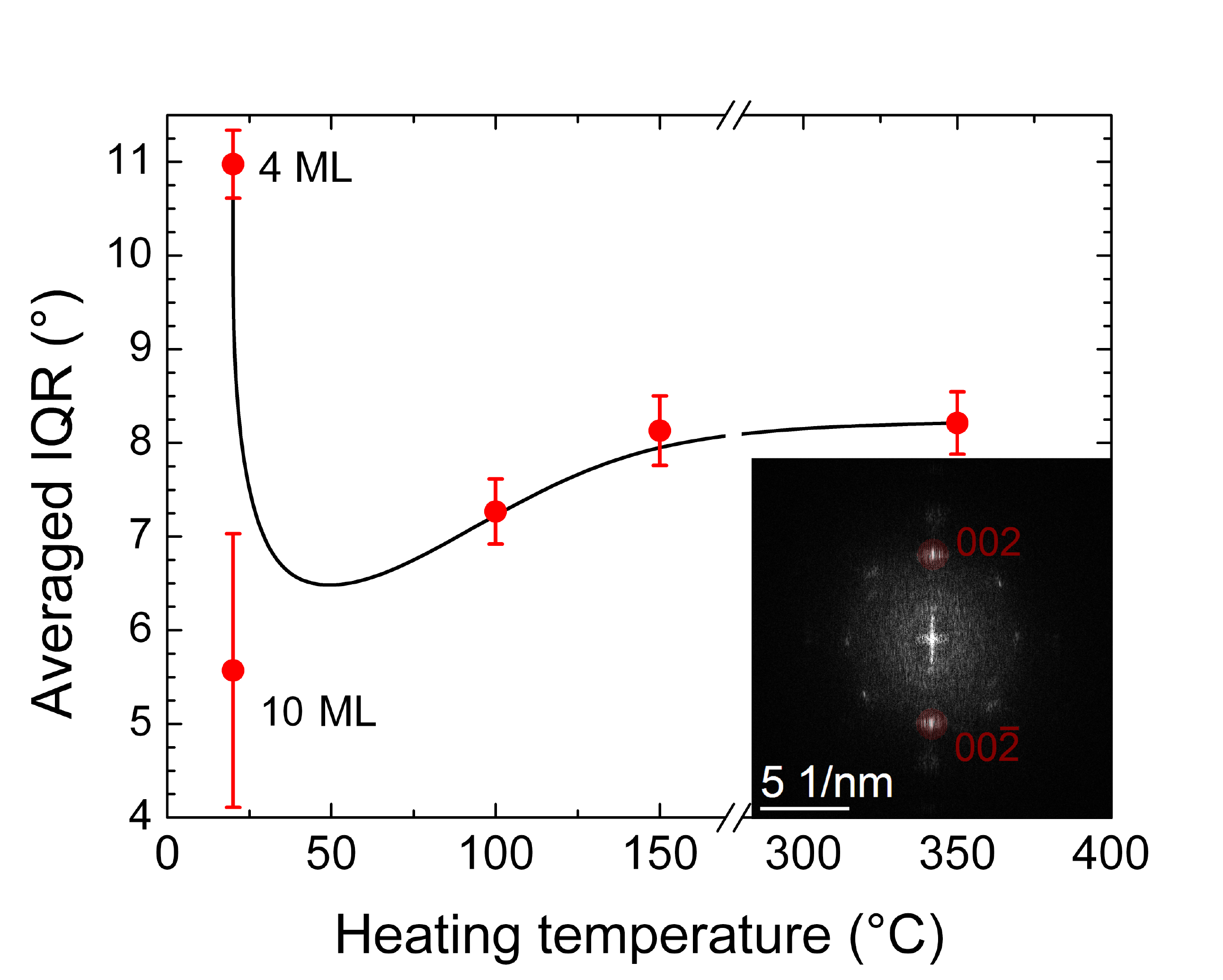}}
\caption{Quantitative analysis of the HR-TEM images of ultra thin MgO barriers at different growth temperatures. The black line is a guide to the eye. The IQR values for room temperature and 100$^\circ$C  were obtained in 4~ML MTJs, 150$^\circ$C  and 350$^\circ$C in 3~ML MTJs.
The inset shows the space frequencies (red) used for Fourier filtering of the 3~ML MgO barrier from Fig. \ref{vergleich_dick_duenn}.}
\label{iqr}
\end{figure}

For example, the high degree of texturing of the 10~ML layer in 
Fig.~\ref{vergleich_dick_duenn} (left), left is estimated as $(5.6~\pm~1.5)^\circ$, whereas the 
lower degree of texturing of the 3~ML layer in Fig.~\ref{vergleich_dick_duenn} (right) 
gives a value of $(6.7~\pm~0.8)^\circ$. 

This allows us to quantify the degree of order and to optimize the barrier growth conditions. 
We applied the method to find the optimal MgO growth temperature. The IQR value is strongly 
decreased with an MgO growth temperature of 100$^\circ$C. 
However, we found no further enhancement for growth temperatures higher than 100$^\circ$C (Fig. 
\ref{iqr}).

\section{Spin-transfer torque in PMA junctions}
First, R(H) and I(U) measurements are carried out to characterize the samples. The 10~ML MgO barriers show a
TMR ratio of up to 270\% and a large magneto-Seebeck effect of up to 50\%. For 3~ML of MgO, a TMR ratio 
of up to 55\% is found.

\begin{figure}[htbp]
\centerline{\includegraphics[width=0.62\linewidth]{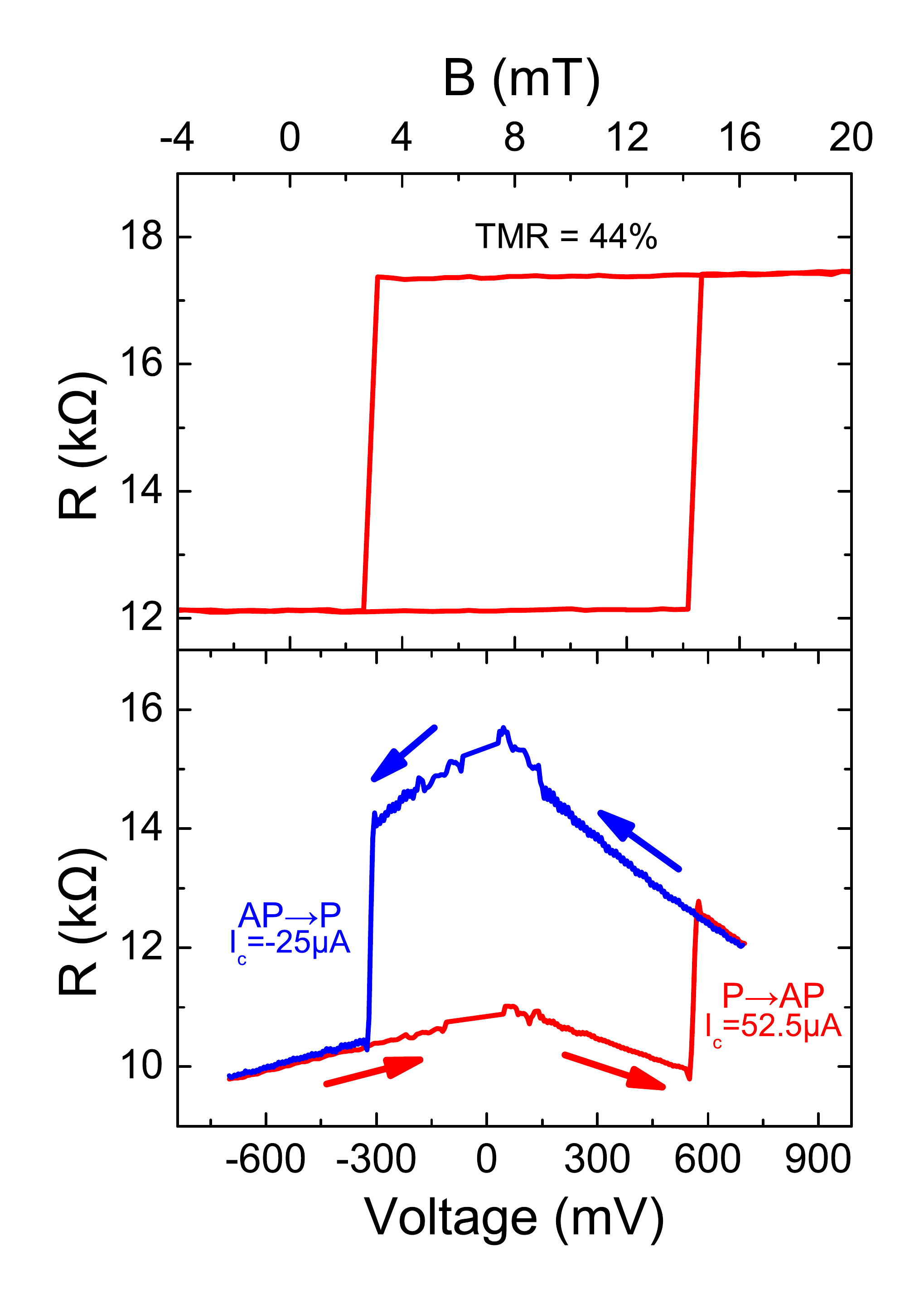}}
\caption{Electrical characterization of an MTJ with 0.8~nm MgO barrier and PMA. top: minor loop, bottom: RV-characteristic with an applied field of 8.6 mT.}
\label{charakterisierung}
\end{figure}

New materials with high perpendicular magnetization anisotropy are important for high 
density storage applications. For example, MTJs with Mn-Ga electrodes are very promising 
due to their high PMA.\cite{Ma2012} Fortunately, PMA has been demonstrated for Co-Fe-B 
films below a critical thickness. Due to the small thickness, which is approximately 
equal to the absorption length of the transferred torque, this results in a reduction of 
the critical switching current $J_c$.\cite{CHzwo,Ikeda2010}
Secondly, $J_c$ is reduced, because the increase in PMA in these junctions 
 lowers the influence of the out-of-plane demagnetizing field which is one of the 
 factors responsible for a high $J_c$.\cite{applphyslett98112507,sun2000} As a 
 consequence, we expect that this reduction of $J_c$ also applies to the thermally 
 driven electron transport through the device and thus enhances the possibility of 
 observing T-STT.
 
Thus, the thickness of both Co-Fe-B layers was reduced to 1.2~nm and 1.0~nm. The resistance is plotted as function of the magnetic field perpendicular to the plane in Fig.~\ref{charakterisierung}. 
The upper viewgraph shows PMA for this thickness range of both magnetic 
layers. The MgO thickness is 4~ML (0.8~nm) in this case with a maximum TMR ratio of 
64\%.For the TMS measurements the MTJ was heated from the top with a diode 
laser (wavelength 640~nm, power up to 150~mW), modulated by a frequency generator at 1.5~kHz. The voltage is then recorded using a lock-in amplifier (see Ref. 5 for more details). A TMS effect of 6\% was 
observed.

The lower graph of Fig.~\ref{charakterisierung} shows the junction resistance as a
function of the applied bias voltage with a magnetic offset field of 8.6~mT. Spin-transfer 
torque switching from the AP to the P state as well as from the P to the AP state can be seen. 
The corresponding critical currents are -25~$\mu$A and 52.5~$\mu$A. Given 
a diameter of 155~nm of the circular junction and using an average critical current of 
38.5~$\mu$A, the average critical current density equals 0.2~MA/cm$^2$. This value is 
much lower than the critical current densities of more than 0.8~MA/cm$^2$ reported by other groups so 
far.\cite{Ikeda2010,Hayakawa2005,Matsumoto2009}


\section{Torque simulations}

We also perform ab initio calculations based on density functional theory. In particular, we use the Korringa-Kohn-Rostoker and the non-equilibrium Green's function method\cite{CHeins,CHzwo} to obtain the thermal spin-transfer torque of the tunnel junction. 

In particular, the torque acting on the atomic layer $i$ is given by
\begin{equation}
\vec{\tau}_i(E)=\frac{1}{\hbar}\Delta_i \hat{M}_i \times \delta \vec{m}_i(E),
\end{equation}
where $\Delta_i$ is the exchange energy, $\vec{M}_i$ is the magnetic moment, and $\hat{M}_i=\vec{M}_i/M_i$. The change in the magnetic moment in each layer $\delta \vec{m}_i$ due to the current is calculated using the non-equilibrium Green's function formalism where the details of our implementation are given in Ref. 23.
To get the energy dependent torque acting on the free layer $\tau^{free}(E)$, we sum up over all atomic layers within the free layer. By integrating over energy we get for the total torque in the free layer
\begin{eqnarray}
\tau^{free}  = \int  ( \tau^{free}_{L \rightarrow R} f_L(E,T_L,\mu)+ \tau^{free}_{R \rightarrow L} f_R(E,T_R,\mu)   ) dE,
\end{eqnarray}
where $f_{L(R)}$ and $T_{L(R)}$ are the occupation function and temperature of the left (right) lead. $\mu$ is the electro-chemical potential and $\tau^{free}_{L \rightarrow R (R \rightarrow L)}$ is the torque acting on the free layer originating from electrons going from left to right (right to left).
The system consists of a tunnel barrier with 3~ML MgO between two ferromagnetic Fe leads with 20~ML and Cu as a reservoir. 

\begin{figure}[htbp]
\centerline{\includegraphics[width=0.65\linewidth]{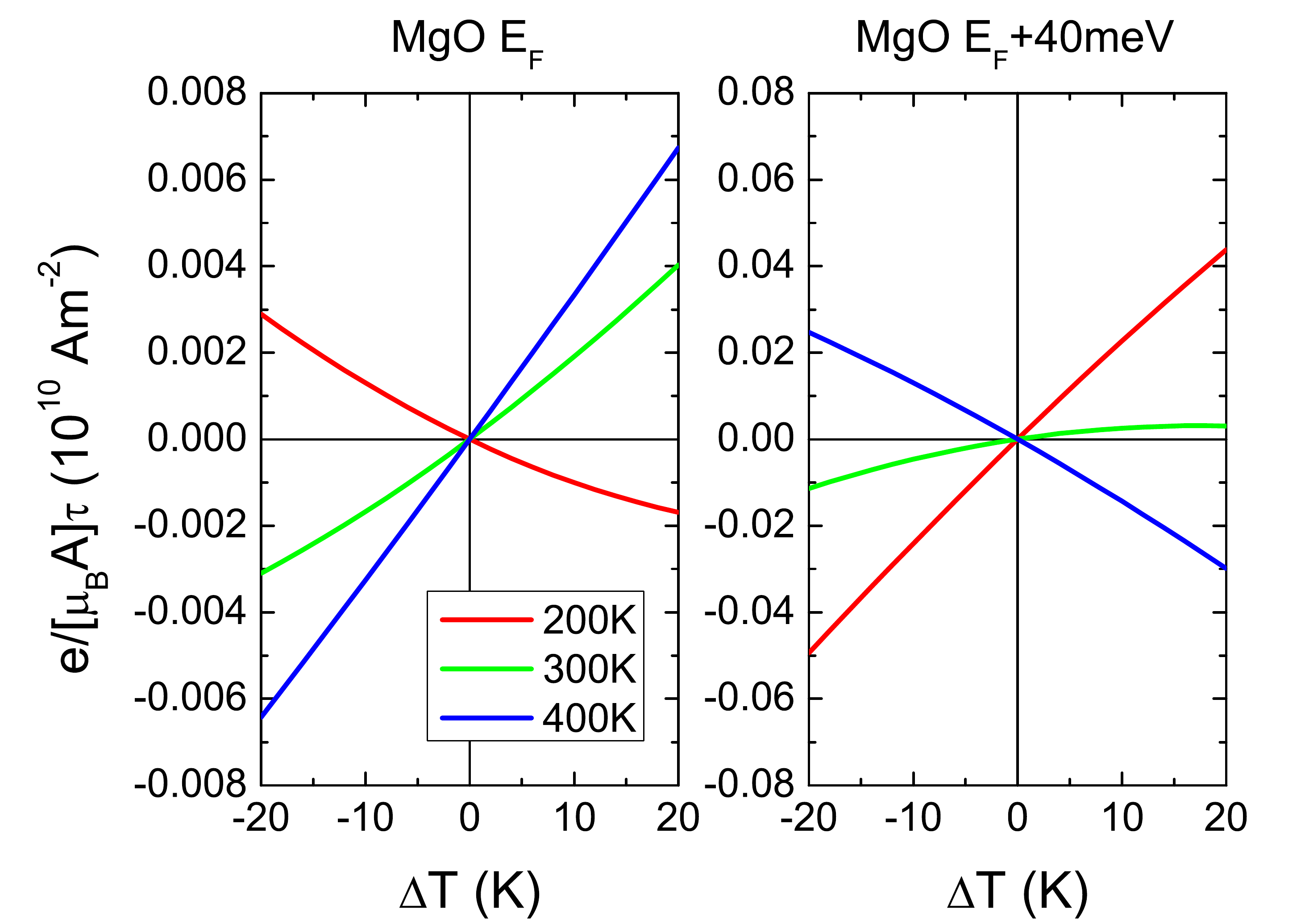}}
\caption{Calculated thermal torque at different lead temperatures for a 3 ML MgO barrier (left). With shifting the CoFe Fermi level by 40 meV (right) the acting torque can be enhanced significantly.}
\label{MgO3dT}
\end{figure}

\section{Temperature simulations}
It remains to show that a temperature gradient of larger than 10~K can be realized across our ultra thin barriers.  
Here, we want to utilize an ultrashort femtosecond laser pulse to generate large temperature gradients. Then, extreme conditions are attained by the very short deposition of the energy within a very thin layer at the top surface of the device.
This allows temperature gradients of approximately 100\,K within a few nanometer right after excitation and thermalization of the electrons. As the time evolves, this extreme temperature gradient will level out. We applied finite element simulations using COMSOL multiphysics numerical solver to estimate the temperature gradients across the MgO barrier that can be achieved in our experimental geometry.

The magnetic tunnel junction is simulated with a diameter of $200\,\mathrm{nm}$ and is 
modeled in a two dimensional, rotational symmetric geometry. The simulated layer stack is as follows: 3$\mu$m Si substrate / 500~nm SiO$_2$ / 10~nm Ta / 2.5~nm 
Co-Fe-B / 0.63~nm MgO / 5.4~nm Co-Fe-B / 5~nm Ta / 3~nm Ru / 36.6~nm Au. According to our 
lithography process, the layers down to the 10\,nm thick Ta layer are 
patterned as a junction pillar which is isolated by $50.4\,\mathrm{nm}$ of SiO$_2$. The 
top strip line consists of $6.3\,\mathrm{nm}$ Cr and $25.2\,\mathrm{nm}$ Au. 
The complete model has a diameter of $50\,\mu\mathrm{m}$ which is 
large enough to cover 
the heating effect of the laser pulse with $w_0 \approx 11\,\mu\mathrm{m}$ (see below).
COMSOL's heat transfer module solves the heat conduction equation in which
the femtosecond laser pulse acts as heat source. In 
addition, we have to know the specific heat $c_p$, the density $\rho$ and the thermal conductivity 
$\kappa$ for the materials involved and these values can be found in Walter et al.\cite{Walter2011}
The heat distribution provided by the laser pulse is modeled by the Lambert-Beer law 
and a scaling factor which includes the material's reflectance $R$ as well as the optical 
penetration depth $\lambda$. The spatial distribution of the pulse energy 
$E_{\mathrm{pulse}}$ is assumed to be Gaussian with the beam waist $w_0$.
The temporal shape of the laser pulse is Lorentzian with the pulse duration $\Gamma$.
The parameter values 
are $E_{\mathrm{pulse}} = 0.2\,\mu\mathrm{J}$, $w_0 = 11\,\mu\mathrm{m}$, $\Gamma = 
50\,\mathrm{fs}$, as measured in the confocal microscope setup used and described 
in Walter et al.\cite{Walter2011} The heat distribution provided by the laser pulse was calculated 
with the optical constants found in the literature for gold\cite{Johnson1972,Palik1985}, 
namely a reflectivity at 800~nm of $R=0.975$ and an optical penetration depth at 800~nm 
of $\lambda=12.7\,\text{nm}$. The thickness of the uppermost gold layer is almost 
twice its optical penetration depth. Therefore, most of the heat will be absorbed in the gold layer, which justifies taking only the optical constants for gold into account. The layers were discretized in a fine mesh with element sizes in the junction layers of 0.2~nm -- 2~nm. 

\begin{figure}[htbp]
\centerline{\includegraphics[width=0.7\linewidth]{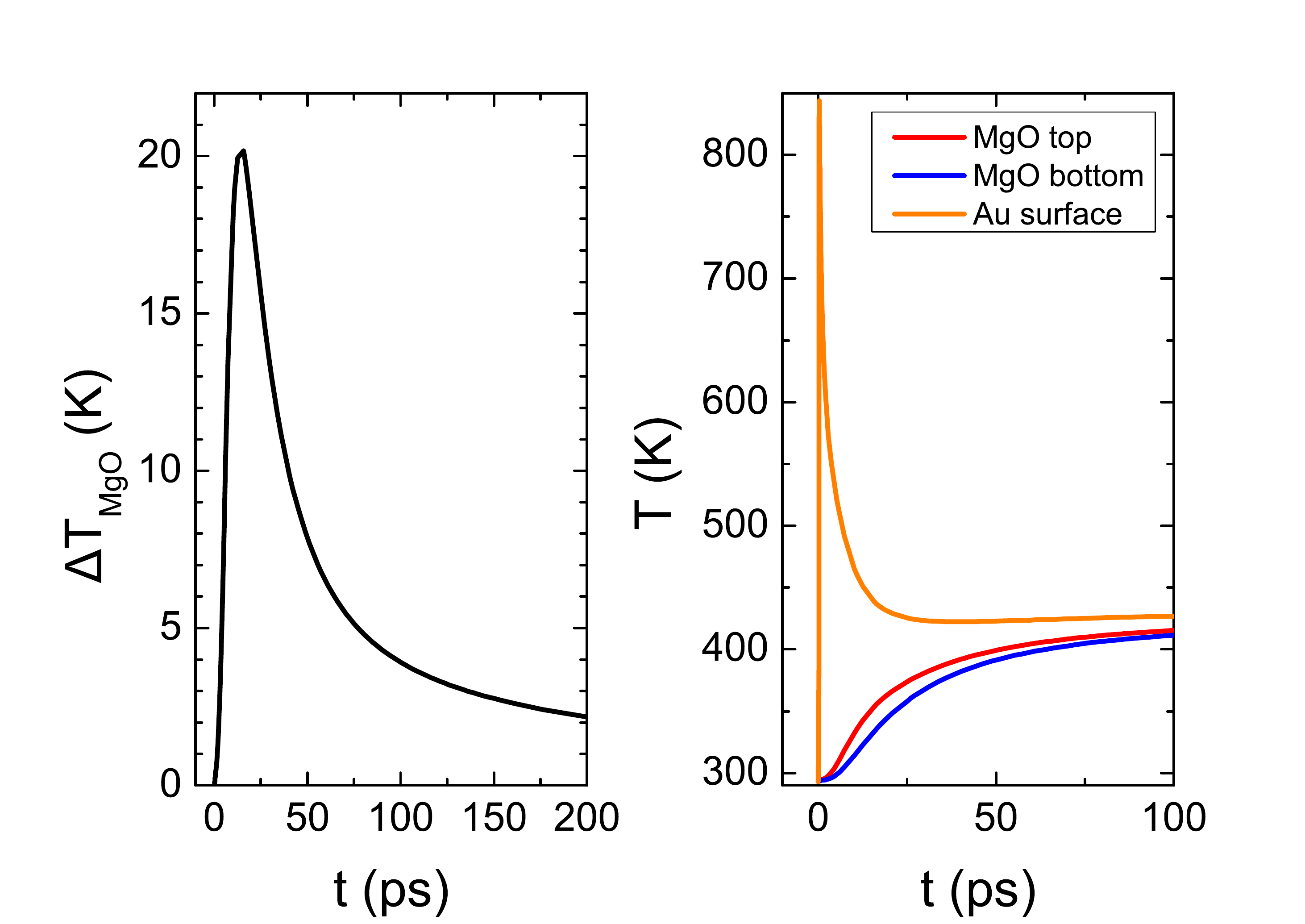}}
\caption{left: Simulated temperature difference across a 0.63~nm MgO tunnel barrier as a 
function of time. right: temperature of the heated gold surface and the two MgO/Co-Fe-B 
interfaces}
\label{fig:temp-sim}
\end{figure}

In the right viewgraph of figure~\ref{fig:temp-sim}, the temperatures for 
the two Co-Fe-B/MgO interfaces and the gold surface in the center of the junction are 
shown. The temperatures as well as its temporal evolution is reasonable for the two 
Co-Fe-B/MgO interfaces. As depicted in Fig.~\ref{fig:temp-sim}, the maximum temperature 
difference across the MgO interface is approximately 20\,K. Temperatures of more than $\Delta T = 
6.5\,\text{K}$ are required for the T-STT switching from the antiparallel into the parallel state according to the calculations of Jia et al.\cite{Jia2011} In total, we achieve a temperature difference of more than a few K for more than a hundred picoseconds for the pulse power and geometry simulated here.


\section{Conclusion}
We demonstrated the fabrication of magnetic tunnel junctions with ultra thin MgO 
barriers and small interface roughness. The samples of 3 to 4\,ML barrier thickness showed high TMR ratios of 55\% to 64\% and spin-transfer torque switching with critical current densities as low as 0.2~MA/cm$^2$.
The texture of the junctions has been investigated by 
high-resolution TEM imaging quantitatively. 
We suggest the average inter-quartile range value of the MgO layer Hough transform as an quantitative indicator 
of the degree of texturing for the junction quality. This allows us to optimize the MgO barrier growth.
The thermal torque has been calculated as a function of the temperature gradient. 
With adjusting the Fermi level via the Co-Fe-B composition, a maximum T-STT effect can be obtained.
Temperature simulations of junctions heated by femtosecond laser pulses revealed large 
temperature gradients in the order of 10~K for around 100~ps. On the base of these parameters, we expect to observe T-STT in these kind of junctions. Thus, magnetic switching by applying only a temperature gradient will be feasible.


\section*{Acknowledgments} \noindent M.M.\ and M.S.\ are supported by German 
Research Foundation (DFG) through SFB~602. C.H., M.M. and A.T. are supported by the DFG 
through SPP~1538 SpinCaT. A.T.\ is supported by the the Ministry of Innovation, Science 
and Research (MIWF) of North Rhine-Westphalia with an independent researcher grant. 
G.R. is supported by the DFG  through grant RE1052/22-1.


\end{document}